\documentclass[apj,twocolumn]{aastex63}
\usepackage[mathscr]{eucal}
\usepackage{amsmath}
\usepackage{systeme}

\shorttitle{Feedback / Cooling Equilibrium Across 10~Gyr of Cluster Growth}
\shortauthors{F. Ruppin \emph{et al.}}

\def\apj{ApJ}%
\def\apjl{ApJ}%
\def\apjs{ApJS}%
\def\aap{A\&A}%
\def\mnras{MNRAS}

\def\chandra{{\it Chandra}}

\newcommand{\MIT}{Kavli Institute for Astrophysics and Space Research, Massachusetts Institute of Technology, 77 Massachusetts Avenue, Cambridge, MA 02139}
\newcommand{\ULyon}{Univ.  Lyon,  Univ.  Claude Bernard Lyon 1, CNRS/IN2P3, IP2I Lyon, F-69622, Villeurbanne, France}
\newcommand{\UMiss}{Department of Physics and Astronomy, University of Missouri—Kansas City, 5110 Rockhill Road, Kansas City, MO, 64110, USA}

\newcommand{\UMont}{Département de Physique, Université de Montréal, C.P. 6128, Succ. Centre-Ville, Montréal, QC H3C 3J7, Canada}

\newcommand{\UDur}{Department of Physics, University of Durham, South Road, Durham DH1 3LE}
\newcommand{\CFA}{Harvard-Smithsonian Center for Astrophysics, 60 Garden Street, Cambridge, MA 02138}
\newcommand{\LLC}{Huntingdon Institute for X-ray Astronomy,  LLC}
\newcommand{\UCin}{Department of Physics, University of Cincinnati, Cincinnati, OH,45221, USA}
\newcommand{\KIPAC}{Kavli Institute for Particle Astrophysics and Cosmology,Department of Physics, Stanford University, Stanford, CA,94305}
\newcommand{\CDur}{Centre for Extragalactic Astronomy, Durham University, South Road, Durham DH1 3LE, UK}
\newcommand{\IDur}{Institute for Computational Cosmology, Durham University, South Road, Durham DH1 3LE, UK}
\newcommand{\UMel}{School of Physics, University of Melbourne, Parkville, VIC 3010, Australia}
\newcommand{\ANL}{High Energy Physics Division, Argonne National Laboratory, 9700 South Cass Avenue, Lemont, IL 60439, USA}
\newcommand{\KChic}{Kavli Institute for Cosmological Physics, University of Chicago, 5640 South Ellis Avenue, Chicago, IL 60637, USA}
\newcommand{\USyd}{School of Science, Western Sydney University, Locked Bag 1797, Penrith, NSW 2751, Australia}
\newcommand{\CSIRO}{CSIRO Astronomy \& Space Science, P.O. Box 76, Epping, NSW 1710, Australia}
\newcommand{\UWis}{Center for Gravitation, Cosmology, and Astrophysics, Department of Physics, University of Wisconsin-Milwaukee, P.O. Box 413, Milwaukee, WI 53201, USA}
\newcommand{\UChic}{Department of Astronomy and Astrophysics, University of Chicago, 5640 South Ellis Avenue, Chicago, IL 60637}

\begin{document}

\title{Redshift Evolution of the Feedback / Cooling Equilibrium in the Core of 48 SPT Galaxy Clusters:\\A Joint \textbf{\emph{Chandra}}\,--\,SPT\,--\,ATCA analysis}

\author[0000-0002-0955-8954]{F.~Ruppin}\affiliation{\MIT}\affiliation{\ULyon}
\author{M. McDonald}\affiliation{\MIT}
\author{J. Hlavacek-Larrondo}\affiliation{\UMont}
\author{M.  Bayliss}\affiliation{\UCin}
\author[0000-0001-7665-5079]{L. E. Bleem}\affiliation{\ANL}\affiliation{\KChic}
\author[0000-0002-2238-2105]{M.  Calzadilla}\affiliation{\MIT}
\author{A.  C.  Edge}\affiliation{\UDur}
\author{M.  D. Filipovi\'c}\affiliation{\USyd}
\author[0000-0003-4175-571X]{B. Floyd}\affiliation{\UMiss}
\author{G.  Garmire}\affiliation{\LLC}
\author[0000-0002-3475-7648]{G.  Khullar}\affiliation{\UChic}\affiliation{\KChic}\affiliation{\MIT}
\author[0000-0001-6506-0293]{K. J.  Kim}\affiliation{\UCin}
\author{R.  Kraft}\affiliation{\CFA}
\author[0000-0003-3266-2001]{G.  Mahler}\affiliation{\CDur}\affiliation{\IDur}
\author{R.  P. Norris}\affiliation{\USyd}\affiliation{\CSIRO}
\author[0000-0003-4609-2791]{A. O'Brien}\affiliation{\UWis}
\author{C. L.  Reichardt}\affiliation{\UMel}
\author{T.  Somboonpanyakul}\affiliation{\KIPAC}
\author{A. A.  Stark}\affiliation{\CFA}
\author{N.  Tothill}\affiliation{\USyd}
%\author{L. E. Bleem}\affiliation{\ANL}\affiliation{\KChic}
%\author{S.~W. Allen}\affiliation{\KStan}\affiliation{\UStan}\affiliation{\SLAC}
%\author{B.~A. Benson}\affiliation{\FLab}\affiliation{\KChic}\affiliation{\UChic}
%\author{M. Calzadilla}\affiliation{\MIT}
%\author{G. Khullar}\affiliation{\UChic}
%\author{B. Floyd}\affiliation{\UMiss}

\correspondingauthor{F. Ruppin}
\email{ruppin@mit.edu}

\begin{abstract}

We analyze the cooling and feedback properties of 48 galaxy clusters at redshifts $0.4 < z < 1.3$ selected from the South Pole Telescope (SPT) catalogs to evolve like the progenitors of massive and well-studied systems at $z{\sim}0$. We estimate the radio power at the brightest cluster galaxy (BCG) location of each cluster from an analysis of Australia Telescope Compact Array (ATCA) data. Assuming that the scaling relation between radio power and active galactic nucleus (AGN) cavity power $P_{\mathrm{cav}}$ observed at low redshift does not evolve with redshift, we use these measurements in order to estimate the expected AGN cavity power in the core of each system. We estimate the X-ray luminosity within the cooling radius $L_{\mathrm{cool}}$ of each cluster from a joint analysis of the available \chandra\ X-ray and SPT Sunyaev-Zel’dovich (SZ) data. This allows us to characterize the redshift evolution of the $P_{\mathrm{cav}} / L_{\mathrm{cool}}$ ratio. When combined with low-redshift results, these constraints enable investigations of the properties of the feedback/cooling cycle across 9~Gyr of cluster growth.  We model the redshift evolution of this ratio measured for cool core clusters by a log-normal distribution $\mathrm{Log}$-$\mathcal{N}(\alpha + \beta z, \sigma^2)$ and constrain the slope of the mean evolution $\beta = -0.05\pm 0.47$. This analysis improves the constraints on the slope of this relation by a factor of two. We find no evidence of redshift evolution of the feedback/cooling equilibrium in these clusters which suggests that the onset of radio-mode feedback took place at an early stage of cluster formation. High values of $P_{\mathrm{cav}} / L_{\mathrm{cool}}$ are found at the BCG location of non-cool core clusters which might suggest that the timescales of the AGN feedback cycle and the cool core / non-cool core transition are different. This work demonstrates that joint analyses of radio, SZ, and X-ray data solidifies the investigation of AGN feedback at high redshifts.

\end{abstract}


\section{Introduction}\label{sec:intro}

\begin{figure*}[t]
\centering
\includegraphics[height=10cm]{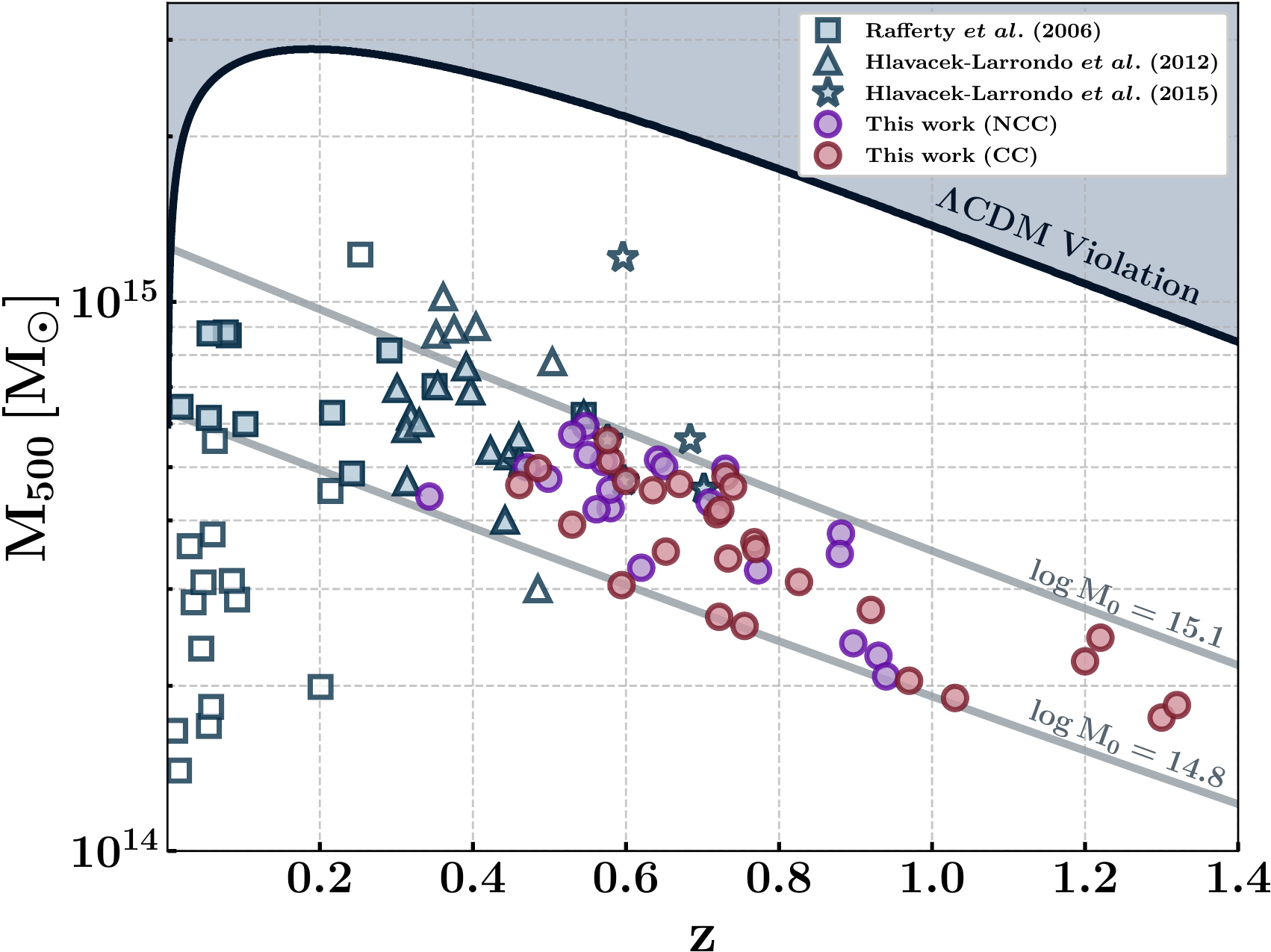}
\caption{{\footnotesize Mass and redshift distribution of the 48 clusters considered in this work (circles).  We sub-divided the sample into cool cores (red) and non-cool core systems (purple). We also show the clusters considered in \cite{raf06} (squares),  \cite{hla12} (triangles), and \cite{hla15} (stars) with significant detections of X-ray cavities.  They enabled the study of the feedback/cooling equilibrium at lower redshifts.  The diagonal lines give the redshift evolution of the mean mass growth obtained by \cite{fak10} for clusters with total mass between $6.3 \times 10^{14}$~M$_{\odot}$ and $1.3 \times 10^{15}$~M$_{\odot}$ at $z=0$.  Clusters that do not pass our progenitor selection cuts (diagonal lines) are shown with empty symbols.  We indicate the 90\% confidence level exclusion limit on the cluster abundance given the considered cosmological model (black line).}}
\label{fig:m_z_plane}
\end{figure*}

% Why is it important to understand the onset of radio mode feedback
Early investigations of the properties of the intracluster medium (ICM) surrounding the brightest cluster galaxy (BCG) of galaxy clusters revealed central cooling times significantly shorter than the age of the universe \citep[\emph{e.g.}][]{fab77,edg92,san06}. Neither massive reservoirs of cold gas, nor the consequent high star formation rates,  have however been observed in the cores of the vast majority of these systems \citep[\emph{e.g.}][]{fab94,fog17,mcd18}. A proposed solution to this cooling-flow problem is that cooling is balanced by non-gravitational processes induced by the supermassive black hole at the center of the BCG that inject energy back into the ICM \citep[\emph{e.g.}][]{voi05,fab06,gas12}. The accretion rate of these active galactic nuclei (AGN) at high redshifts is very close to the Eddington limit, which leads to radiative quasar-mode feedback \citep[\emph{e.g.}][]{fab12}. On the other hand, most AGN observed in the core of nearby clusters present much lower accretion rates and induce mechanical radio-mode feedback in the form of powerful jets that carve cavities into the ICM \citep[\emph{e.g.}][]{ran11}. Unveiling the onset of radio-mode feedback at high redshift is essential to understand this transition in the accretion rate of central AGN and its impact on cluster formation.\\
% Methode and results from previous studies
\indent To this end, several X-ray analyses have been conducted on samples of giant elliptical galaxies and clusters in order to detect and characterize cavities and study the equilibrium between cooling and feedback \citep[\emph{e.g.}][]{raf06,nul09,hla12,hla15,hla20}.  For example, \cite{hla15} studied 83 clusters at $0.4 < z < 1.2$ selected from the the South Pole Telescope (SPT) catalog \citep{ble15} with available \chandra\ data in order to detect and characterize X-ray cavities around the BCG. These studies, however,  only enabled significant detections of cavities at redshifts $z < 0.8$ because of the strong redshift dimming of the X-ray surface brightness. Out of the 83 clusters considered by \cite{hla15}, only 6 presented convincing cavities around the BCG (see Fig. ~\ref{fig:m_z_plane}). As detecting X-ray cavities and characterizing the central cooling properties of clusters at high redshift is extremely challenging with current or planned near-term X-ray observatories, it is essential to propose new methods in order to keep pushing the investigation of AGN feedback to higher redshifts.\\ 
% What we do in this work
\indent In this work, we present a joint analysis of radio, Sunyaev-Zel’dovich (SZ), and X-ray observations realized with the Australia Telescope Compact Array (ATCA),  SPT, and \chandra, respectively. We use the ATCA data in order to estimate the AGN jet power at the BCG location of 48 SPT clusters at $0.4 < z < 1.3$ selected to evolve like the progenitors of well-studied systems at $z{\sim}0$. We further estimate the X-ray luminosity within the cooling radius of these clusters from the joint analysis of the \chandra\ and SPT data in order to characterise the redshift evolution of the feedback / cooling balance during cluster growth.\\ 
% Paper organisation and cosmology
\indent In \textsection \ref{sec:sample_data} we summarize the cluster selection procedure as well as the data used in this paper. In \textsection \ref{sec:pcav} we present how we estimate the cavity power in the core of each cluster. The measurement of the associated cooling luminosity is described in \textsection \ref{sec:lcool} and the characterization of the redshift evolution of the cooling / feedback balance is presented in \textsection \ref{sec:results}. In \textsection \ref{sec:discussion} we discuss the implications of this study for AGN feedback at high redshift before summarizing our work in \textsection \ref{sec:conclu}. Throughout this paper, we consider a flat $\Lambda$CDM cosmology with $\Omega_m = 0.3$, $\Omega_{\Lambda} = 0.7$, and $H_0 = 70~\mathrm{km\,s^{-1}\,Mpc^{-1}}$.

\section{Cluster Selection and Data Set}\label{sec:sample_data}

\begin{figure*}[t]
\centering
\includegraphics[height=8cm]{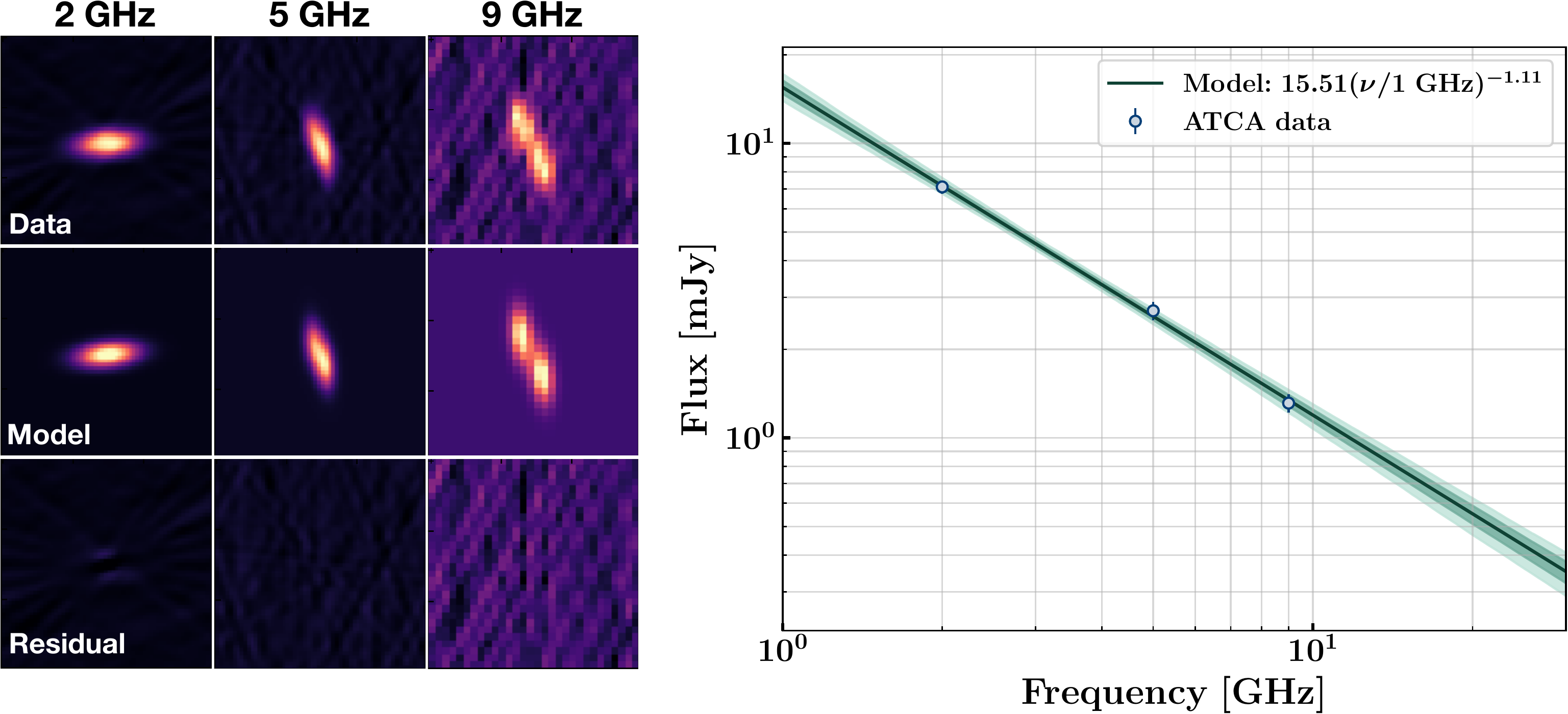}
\caption{{\footnotesize \textbf{Left:} From top to bottom, ATCA data, map model, and residual of the radio AGN detected at the center of the SPT-CLJ0307-6225 cluster at 2, 5, and 9~GHz,  from left to right, respectively. \textbf{Right:} Flux densities associated with the models shown in the left panel (blue points) along with the best-fit SED model (green line) and its associated $1\sigma$ and $2\sigma$ confidence regions (green areas).}}
\label{fig:radio_data}
\end{figure*}

% Reminder about the progenitor-selected sample
The cluster selection procedure is detailed in \cite{rup21}; we briefly summarize it here. We use the analytic formula for the mean mass growth rate of haloes as a function of redshift obtained by \cite{fak10} in order to select clusters from the SPT catalogs defined in \cite{ble15} and \cite{hua19}. The selected clusters are the progenitors of halos with a $z=0$ mass enclosed between M$_{500} = 6.3 \times 10^{14}$~M$_{\odot}$\footnote{M$_{500}$ is defined as the mass enclosed within a sphere with a mean mass density equal to 500 times the critical density of the Universe at the cluster redshift.} and M$_{500} = 1.3 \times 10^{15}$~M$_{\odot}$ at $z=0$.\\
% Description of the Chandra observations
\indent Among all SPT clusters satisfying this condition, 73 have been observed by \chandra\ as part of four dedicated projects. The \chandra\ X-ray Visionary Project (XVP; PI: B. Benson) described in \cite{mcd13} enabled obtaining ${\sim}1300$~counts in the 0.7-2~keV band for 49 clusters during \chandra\ cycles 12 and 13.  Three of these clusters have been further observed with \chandra\ thanks to another large program (PI: J.  Hlavacek-Larrondo).  A third large program (PI: M. McDonald) targeted 18 clusters at $z>0.7$ and allowed us to obtain an average of 180 counts per cluster \citep{rup21}. The latest program (PI: F. Ruppin), targeting 7 clusters at $z > 0.9$ is currently on-going but the observations of 6 of these clusters are completed.\\
% Description of the ATCA observations
\indent In this paper, we aim to study the evolution of the cooling and feedback balance in the cores of clusters lying along a common evolutionary track. To this end, we further consider radio observations realized with ATCA for a sub-sample of these clusters during three separate observing runs. A single map of the whole $100~\mathrm{deg}^2$ SPTpol footprint \citep{hua19} was realized in May 2013 in the 6A configuration at 2~GHz with a root mean square (rms) noise varying between 60 and 120~$\mu$Jy/beam across the map \citep{obr16}. Targeted observations of XVP clusters were made in January 2015 at 2~GHz with an rms noise varying between 28 and 55~$\mu$Jy/beam. Some of these clusters were further followed-up at 5 and 9~GHz in August 2016 if a strong detection was made at 2~GHz. We reach an rms noise varying between 30 and 67~$\mu$Jy/beam at 5~GHz and between 19 and 77~$\mu$Jy/beam at 9~GHz. All observations have been reduced using the 05/21/2015 release of the \texttt{MIRIAD} software \citep{sau95}.\\
% Sample selection: exclusion of clusters considering BCG location
\indent Among the 73 SPT clusters satisfying our progenitor selection, 48 have available ATCA data.  If a significant radio source is observed in the ATCA data of a given cluster, we make sure that it is located within $5$ arcsec of the BCG to exclude potential foreground or background contamination. We estimate the BCG location in these clusters from a visual inspection of available optical and IR imaging from \cite{ble15} and \cite{hua19} based on galaxy size and brightness.  We use the X-ray peak position as extra information to solve cases in which multiple BCGs could be identified in the same cluster. The mass-redshift distribution of this sample of 48 clusters is shown in Fig.~\ref{fig:m_z_plane} (circles) along with samples from previous studies. This sample allows us to study the cooling / feedback equilibrium in a redshift range that was previously unexplored ($0.7 < z < 1.3$) while overlapping with samples that have been characterized by \cite{raf06,hla12,hla15}. This enables validation of our methodology that does not rely on X-ray cavity detection in contrast to these previous surveys. We only consider the cool core clusters (red points in Fig.~\ref{fig:m_z_plane}) in the following in order to match the cluster properties of these low redshift samples as much as possible.  The cool core / non-cool core discrimination is performed by the joint analysis of the \chandra\ and SPT observations of the selected clusters following the procedure described in \cite{rup21} (see \textsection \ref{sec:lcool}). We will discuss the results obtained for the non-cool core clusters (purple points) in \textsection \ref{subsec:non_cc_clusters}.

\section{Estimation of Cavity Power}\label{sec:pcav}

% Difficulty to detect cavities at high redshift on large samples --> scaling relation
As detecting X-ray cavities at high redshift with current facilities is extremely challenging, we propose to rely on the scaling relation between AGN jet power and radio power at 1.4~GHz calibrated by \cite{cav10}. Instead of measuring the properties of X-ray cavities to estimate the power of the AGN jets that carved them, we assume that the \cite{cav10} scaling relation does not evolve with redshift and we measure the AGN radio power from the ATCA observations of each cluster considered in this work to infer the corresponding jet power. We stress that the \cite{cav10} scaling relation considers the total radio power measured at 1.4~GHz without discriminating between synchrotron emission from the mini-halo and the AGN lobes. We can therefore use the total radio power estimated with ATCA at 1.4~GHz in order to infer the associated jet power.\\
% Method to measure the radio flux with ATCA data --> source model
\indent We model the ATCA radio data using a sum of 2D Gaussian functions with a position angle and minor and major axes lengths fixed to the ones of the PSF model in each frequency band. If an AGN is detected within $<5"$ of the BCG location in the considered ATCA map, we perform a Markov Chain Monte Carlo (MCMC) analysis in order to estimate the best-fit values of the amplitude and sky-position of each Gaussian function considered in our model. We iterate this analysis with an increasing number of Gaussian functions until the best-fit $\chi^2$ value does not decrease significantly given the RMS noise in the data. The AGN flux density is obtained by integrating the signal in the best-fit model and the corresponding uncertainty is estimated by sampling the posterior distribution of all model parameters. If the AGN detection is not associated with the BCG or no radio AGN is detected in the cluster region, we estimate an upper limit on the AGN flux density based on the ATCA RMS noise measured in a region empty of radio signal within a 5~arcmin radius from the BCG location.\\
% Results on flux modeling
\indent We show the results obtained with the ATCA data of SPT-CLJ0307-6225 in Fig.~\ref{fig:radio_data}. We use it as a representative example of a cluster with a significant AGN detection. The signal in most maps at 2~GHz is well modelled with a single 2D Gaussian function as shown in the left column of the left panel. Indeed, the ATCA angular resolution at this frequency is often too small to resolve the AGN signal. Our model is flexible enough to subtract all the significant signal observed in the data as shown in the residuals shown in the bottom row.\\

\begin{figure*}[t]
\centering
\includegraphics[height=10cm]{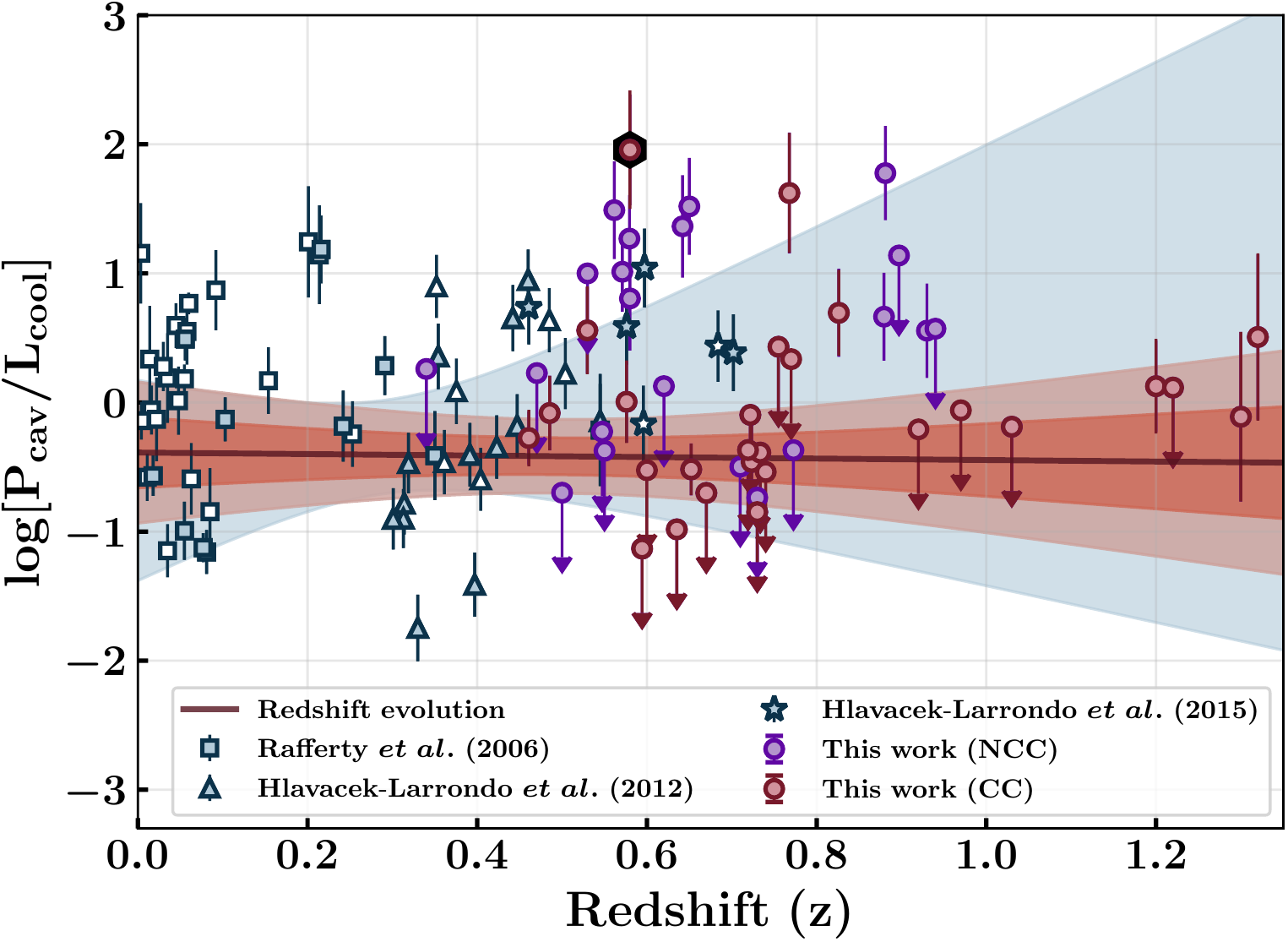}
\caption{{\footnotesize Ratio between the AGN mechanical power and the X-ray luminosity within the cooling radius at 7.7~Gyr in logarithmic scale, for different samples, as a function of redshift. The blue area corresponds to the $2\sigma$ confidence region of the power law fit to the \cite{raf06} and \cite{hla12} samples. The red line and its associated $1\sigma$ and $2\sigma$ confidence regions shows the best-fit redshift evolution of the $P_{\mathrm{cav}} / L_{\mathrm{cool}}$ ratio including our sample of 27 cool core clusters (red points). We also show the results obtained for the non-cool core clusters (purple points) for our discussion in \textsection \ref{subsec:non_cc_clusters}.  Neither the non-cool core points nor the \cite{hla15} data are included in the power law fits.  We highlight SPT-CLJ2245-6206 with a black background hexagon. }}
\label{fig:redshift_evol}
\end{figure*}

% Method to estimate radio power at 1.4 GHz from SED fit
Among the 48 clusters considered in this work, 11 have been observed at 2, 5, and 9~GHz with ATCA. We use the flux densities estimated in each band in order to fit the spectral energy distribution (SED) of the AGN at the core of these systems. The SED is modelled as a power law with free amplitude at 1~GHz and spectral index.  The choice for this model is motivated by previous studies of AGN radio emission in clusters such as \cite{kok17} who find a very good agreement between this model and the data obtained in the $[0.1-10]$~GHz range.  We show the best-fit SED model of the AGN detected in SPT-CLJ0307-6225 in the right panel of Fig.~\ref{fig:radio_data} along with its associated $1\sigma$ and $2\sigma$ confidence regions. The mean spectral index measured for this sample of 11 clusters is $\alpha = 1.12 \pm 0.06$.  We have checked that this sub-sample of clusters is not biased towards particular core properties. Among these 11 clusters, 5 are classified as cool cores in \citep{rup21} and 6 as non-cool cores. Therefore, we use the mean spectral index estimated with this sample to model the SED of all other 37 clusters with only one ATCA measurement at 2~GHz. Knowing the redshift of every cluster in our sample, we use these SED models to estimate the rest frame flux density at 1.4~GHz, $S_{1.4~\mathrm{GHz}}$. The radio power at this frequency is then given by:
\begin{equation}
P_{1.4~\mathrm{GHz}} = 4\pi D_L^2(1+z)^{\alpha - 1} S_{1.4~\mathrm{GHz}} \times (1.4~\mathrm{GHz})
\end{equation}
where $D_L$ is the cluster luminosity distance. The uncertainty on the radio power is obtained by sampling the error bars on $S_{1.4~\mathrm{GHz}}$, which take into account both the measurement error in the ATCA bands and the uncertainty on the spectral index, as well as the error on the redshift that accounts for ${\sim}5$\% of the total error budget through the $D_L^2$ factor. The associated AGN cavity power $P_{\mathrm{cav}}$ is further deduced from the \cite{cav10} scaling relation:
\begin{equation}
\mathrm{log}\,P_{\mathrm{cav}} = 0.75(\pm 0.14) \mathrm{log}\,P_{1.4~\mathrm{GHz}} + 1.91 (\pm 0.18)
\end{equation}
with an intrinsic scatter $\sigma = 0.78$~dex and a correlation coefficient between the slope and intercept of 0.72. We estimate the uncertainty on the cavity power by sampling the error bars on $P_{1.4~\mathrm{GHz}}$, the covariance matrix associated with the slope and intercept of the \cite{cav10} scaling relation as well as its intrinsic scatter. Our results are summarized in Table~\ref{tab:all_res} and \ref{tab:all_res_NCC}.

\section{X-ray Luminosity within Cooling Radius}\label{sec:lcool}

\begin{table*}
\caption{{\footnotesize Best-fit values and associated uncertainties for the three parameters defining the power law model fitted to the $P_{\mathrm{cav}} / L_{\mathrm{cool}}$ ratios estimated in cool core clusters that satisfy our selection criteria. Results are shown with (bottom row) and without (top row) adding our constraints to the ones obtained in previous studies.}}
\renewcommand{\arraystretch}{1.05}
\begin{tabular}{cccc}
\hline
\hline
Data & $\alpha$ & $\beta$ & $\sigma^2$  \\
\hline
Rafferty \emph{et al. } 2006 + Hlavacek-Larrondo \emph{et al. } 2012 & $-0.60\pm 0.36$ & $0.90\pm 1.09$ & $0.61\pm 0.24$ \\
Rafferty \emph{et al. } 2006 + Hlavacek-Larrondo \emph{et al. } 2012 + This work & $-0.39\pm 0.26$ & $-0.05 \pm 0.47$ & $0.74\pm 0.24$\\
\hline
\end{tabular}
\label{tab:res_fits}
\end{table*}

% Data analysis up to spectroscopic analysis
The X-ray data reduction is made using the Chandra Interactive Analysis of Observations (CIAO) software v4.13 along with the calibration database (CALDB) v4.9.5\footnote{\url{https://cxc.cfa.harvard.edu/ciao/}}. After reprocessing the level 1 event files and removing flares from lightcurves \citep{mar01}, we identify point sources with wavelet filters \citep{vik98} and mask them to produce a cleaned event file. We further extract the X-ray surface brightness profile of each cluster in the 0.7-2.0~keV band using the same binning definition considered in \cite{mcd17} and \cite{rup21}. The center that we consider to extract the X-ray surface brightness profile is the BCG location found with the available optical / IR data. The surface brightness profiles are vignetting-corrected using the exposure map estimated in the same energy band.\\
% Two cases depending on S/N --> X-ray spectrum or joint X-ray / SZ
\indent The analysis procedure used to estimate the ICM thermodynamic profiles of each cluster depends on the quality of the \chandra\ observations.  The 36 clusters observed in the context of the XVP program can be fully processed with an X-ray only analysis as we have enough X-ray counts to constrain the ICM temperature from the analysis of their X-ray spectrum. For these clusters, we extract spectra at different angular distances from the deprojection center requiring at least 500 counts in the 0.7-7.0~keV band. We subtract the particle background using stowed background files scaled to the number of counts observed in the 9-12~keV band. We repeat the same procedure in regions of the ACIS-I chips free from cluster emission in order to estimate the astrophysical background spectrum. We jointly fit the cluster and background spectra using a single-temperature plasma \citep[\texttt{APEC};][]{smi01} model combined with a soft X-ray Galactic background (\texttt{APEC}, $k_BT_X = 0.18~\mathrm{keV}$, $Z = Z_{\odot}$, $z=0$), a hard X-ray cosmic spectrum \texttt{BREMSS} ($k_BT_X = 40~\mathrm{keV}$), and a Galactic absorption model (\texttt{PHABS}). The Galactic column density is fixed to the value given by \cite{kal05}. We fix the cluster redshift to the SPT catalog value and the ICM metallicity to $Z = 0.3 Z_{\odot}$ \citep{man20}. The resulting X-ray temperature measurement allows us to estimate the ICM emission measure profile. We apply the procedure detailed in \cite{rup21} in order to estimate the ICM density profile from a Bayesian forward fit of the emission measure profile based on a Vikhlinin parametric model \citep[VPM;][]{vik06}.\\

The 12 high-redshift clusters in our sample presenting only an average of $180$~counts due to ICM emission over the entire image have been analysed using the joint X-ray/SZ analysis detailed in \cite{rup21}. We jointly fit the \chandra\ surface brightness profile and the SPT integrated Compton parameter using a VPM model for the ICM density and a generalized Navarro-Frenk-White model \citep[gNFW;][]{nag07} for the ICM pressure. This procedure allows us to bypass the analysis of the X-ray spectrum of these clusters as they do not present enough counts to enable measuring reliable X-ray temperatures. Following \cite{hud10}, we classify a cluster as a cool core (CC) if the central ICM density measured at $10$~kpc is such that $n_{e,0} > 1.5 \times 10^{-2}~\mathrm{cm}^{-3}$.\\
% Combination of density and pressure profiles --> cooling time
\indent We finally estimate the isochoric cooling time profile:
\begin{equation}
t_{cool}(r) = \frac{3}{2} \frac{(n_e+n_p) \, k_B T_e}{n_e n_p \, \Lambda(T_e,Z)}
\end{equation}
where $n_p = n_e / 1.199$ is the ICM proton density assuming the ionization fraction of a fully ionized plasma with an abundance of $0.3 Z_{\odot}$ \citep{and89}. The ICM temperature profile $k_BT_e$ is either the one obtained from the deprojection of the \chandra\ spectra of the XVP clusters or the one obtained from the combination of the ICM electron density ($n_e$) and pressure ($P_e$) profiles: $k_BT_e = P_e / n_e$. We use the cooling function estimated by \cite{sut93} for an optically-thin plasma with a $0.3Z_{\odot}$ metallicity to compute $\Lambda(T_e,Z)$.\\
% Estimation of cooling radius at 7.7 Gyr to match previous studies
\indent We estimate the cooling radius of each cluster as the radius enclosing a plasma with a cooling time lower than 7.7~Gyr. We use this threshold to enable comparing our results to the ones obtained in previous studies \citep{raf06,hla12,hla15} that use this definition of the cooling radius. We estimate the X-ray luminosity between 0.2 and 100~keV within the cooling radius using the CIAO tool \texttt{modelflux}. The error bars on the estimated luminosity are obtained by varying the size of the the cooling radius to account for the uncertainty on the cooling time profile as well as by sampling the uncertainty on the ICM temperature within the cooling radius. Our results are summarized in Table~\ref{tab:all_res} and \ref{tab:all_res_NCC}. 

\section{Evidence for Constant Feedback to Cooling Ratio with Redshift}\label{sec:results}

\begin{figure*}[t]
\centering
\includegraphics[height=11cm]{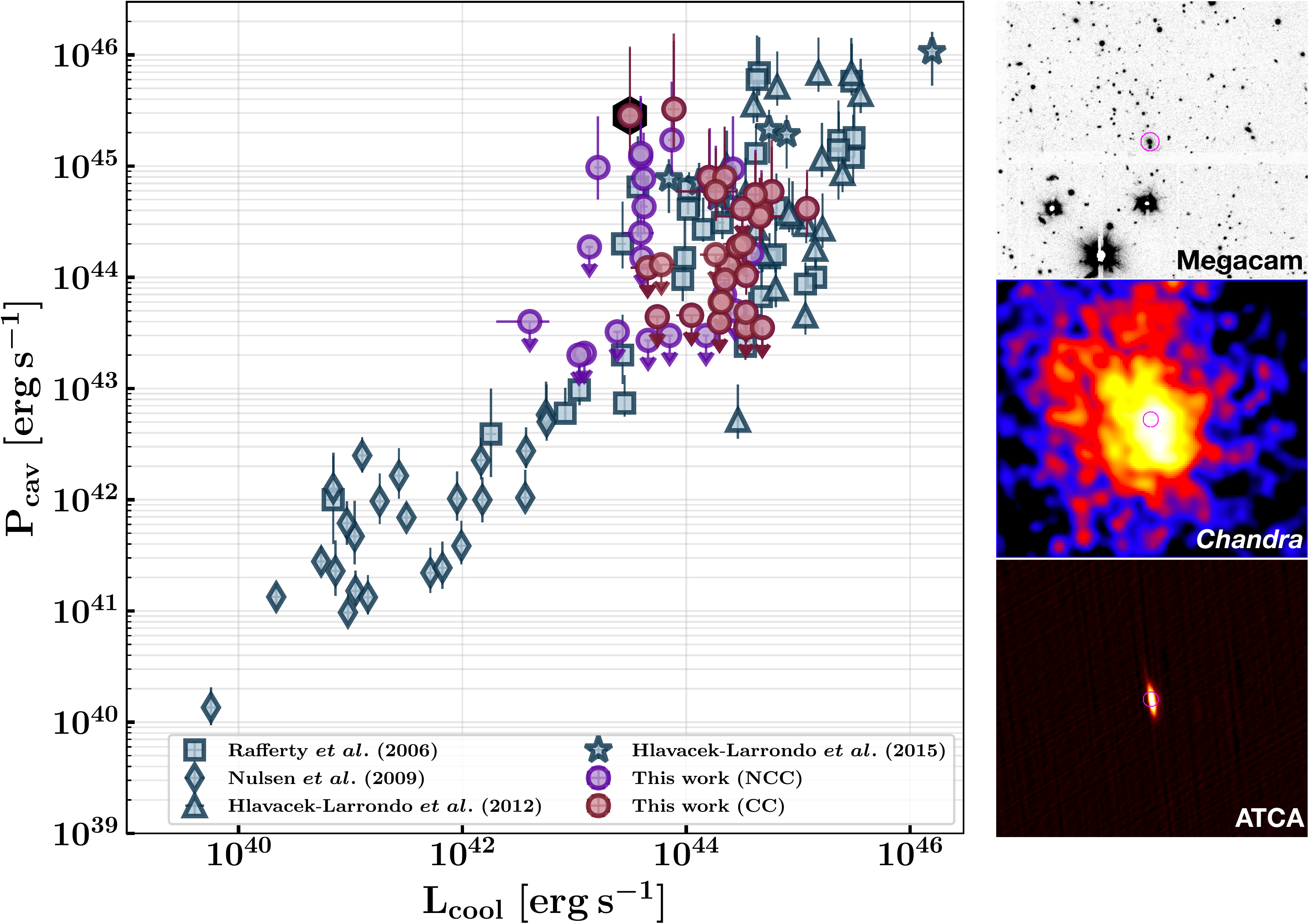}
\caption{{\footnotesize \textbf{Left:} Comparison between the AGN cavity power ($P_{\mathrm{cav}}$) and the X-ray luminosity within the cooling radius at 7.7~Gyr ($L_{\mathrm{cool}}$) of the 48 clusters considered in this work along with results from previous studies.  We split our sample between cool cores (red) and non-cool core systems (purple). We highlight SPT-CLJ2245-6206 with a black background hexagon. \textbf{Right:} Magellan/Megacam $r$ band (top), \chandra\ X-ray (middle), and ATCA radio (bottom) images of the non-cool core cluster SPT-CLJ0542-4100, chosen to be a representative example of this type of clusters in our sample. Each map width is about 2.5~arcmin. The location of the BCG is highlighted with a magenta circle.}}
\label{fig:high_Pc_Lc}
\end{figure*}

% Description of Fig. 3
We use our estimates of the AGN cavity power $P_{\mathrm{cav}}$ (\textsection \ref{sec:pcav}) and X-ray luminosity inside the cooling radius $L_{\mathrm{cool}}$ (\textsection \ref{sec:lcool}) to study the redshift evolution of the $P_{\mathrm{cav}} / L_{\mathrm{cool}}$ ratio in our sample of CC clusters. The results are presented as red points in Fig.~\ref{fig:redshift_evol}. Among the 27 CC clusters in our sample, 15 do not display any significant radio signal at the BCG location in the ATCA data and are presented as upper limits. All but two clusters in our CC sample verify $P_{\mathrm{cav}} / L_{\mathrm{cool}} < 10$, in agreement with results obtained by \cite{raf06,hla12,hla15} at low redshift (blue symbols). The cluster SPT-CLJ0528-5300 at $z = 0.77$ has already been studied in detail by \cite{cal19} who find a ratio $P_{\mathrm{cav}} / L_{\mathrm{cool}} \simeq 63$ in agreement with our estimate. The case of SPT-CLJ2245-6206 at $z = 0.58$ with $P_{\mathrm{cav}} / L_{\mathrm{cool}} \simeq 91$ will be discussed in \textsection \ref{subsec:highest_Pcav_Lcool} along with the results obtained for non-cool core clusters (purple points) in \textsection \ref{subsec:non_cc_clusters}.\\
% Method to fit redshift evolution
\indent We use the Bayesian linear regression package \texttt{LinMix} \citep{kel07} in order to fit the redshift evolution of the $P_{\mathrm{cav}} / L_{\mathrm{cool}}$ ratio, taking upper limits into account. We model the redshift evolution of $\mathrm{log}[P_{\mathrm{cav}} / L_{\mathrm{cool}}]$ as $\mathcal{N}(\alpha + \beta z, \sigma^2)$, where $\alpha$ and $\beta$ are respectively the intercept and slope of the power law defining the mean of the Gaussian distribution $\mathcal{N}$ with intrinsic scatter $\sigma$. The clusters from \cite{raf06,hla12} that satisfy our progenitor selection (filled symbols in Fig.~\ref{fig:redshift_evol}) and our constraints on CC clusters are jointly fit to obtain the dark line in Fig.~\ref{fig:redshift_evol}.  The results of the fits are given in Table~\ref{tab:res_fits} with and without including our constraints in the analysis. The dark and light orange regions show the $1\sigma$ and $2\sigma$ confidence intervals associated with the mean of the distribution.  The blue shaded region corresponds to the $2\sigma$ confidence interval associated with the regression realised by considering the \cite{raf06,hla12} data points only (filled squares and triangles in Fig.~\ref{fig:redshift_evol}).  We note that among the 6 clusters with convincing cavity detections in \cite{hla15}, only 4 satisfy our progenitor selection cuts and they are all located at the high mass end (see Fig.~\ref{fig:m_z_plane}).  Less massive SPT clusters studied by \cite{hla15} in the same redshift range (\emph{i.e.} $0.4 < z < 0.6$) might have shown convincing X-ray cavities with longer exposures.  We therefore chose to exclude these 4 data points from the fit of the redshift evolution of $P_{\mathrm{cav}} / L_{\mathrm{cool}}$ in order to minimize selection effects.\\
% Improvement of the constraints --> intrinsic scatter
\indent We find that including our results reduces the uncertainty on the slope $\beta$ by a factor of 2.3, from $\beta = 0.90\pm 1.09$ to $\beta = -0.05 \pm 0.47$. The high redshift anchor brought by our study enables reaching a regime where the uncertainty on $\beta$ is limited by the intrinsic scatter of the distribution. The measured slope is compatible with an absence of redshift evolution of the feedback to cooling ratio up to $z{\sim}1.5$. 

\section{Discussion and Implications for AGN Feedback}\label{sec:discussion}

\subsection{Onset of radio-mode feedback}\label{subsec:onset}

\begin{figure*}[t]
\centering
\includegraphics[height=5.7cm]{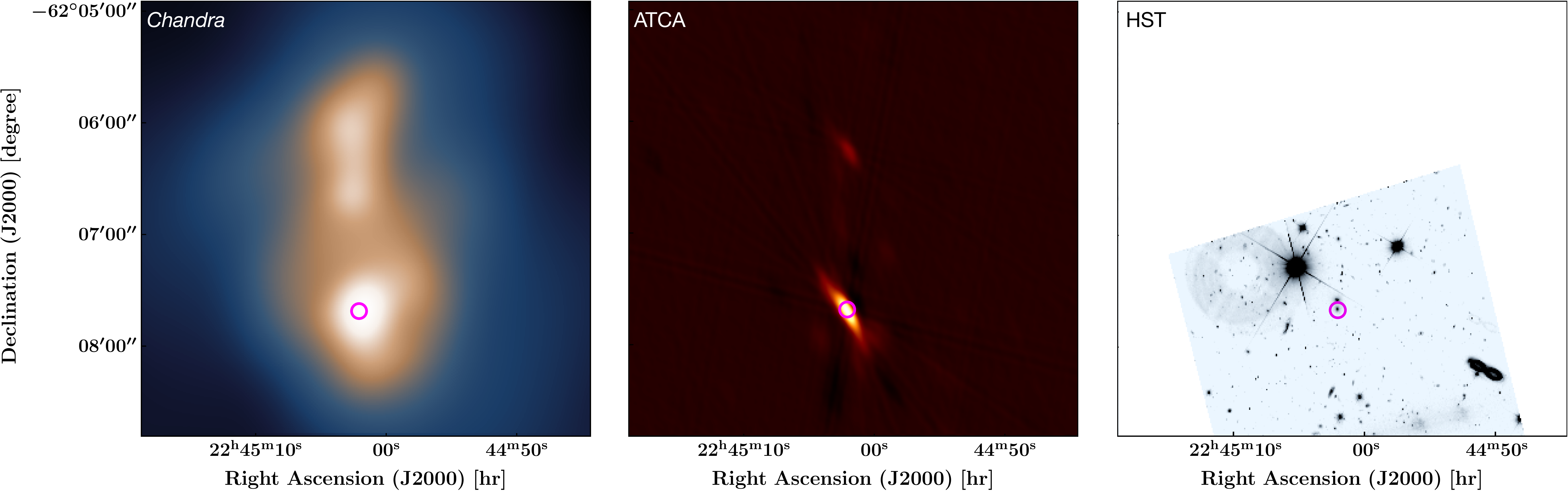}
\caption{{\footnotesize \chandra\ X-ray (left),  ATCA radio (middle), and HST/WFC3-F200LP (right), images of the cool core cluster SPT-CLJ2245-6206.  Each map width is 4~arcmin. The location of the BCG is highlighted with a magenta circle.}}
\label{fig:spt_cl_2245}
\end{figure*}

% Description of Fig. 4 --> no difference with previous samples (probably valid scaling relation)
First, we note that the distribution of CC clusters in the $P_{\mathrm{cav}}—L_{\mathrm{cool}}$ plane shown in Fig.~\ref{fig:high_Pc_Lc} agrees with previous samples at lower redshift. This suggests that the $P_{1.4~\mathrm{GHz}}—P_{\mathrm{cav}}$ scaling relation calibrated by \cite{cav10} does not significantly evolve with redshift.\\
% Radio-mode feedback has been there for a long time
\indent The results presented in \textsection \ref{sec:results} show that the equilibrium between the power generated by AGN mechanical feedback and the cooling of the hot X-ray emitting phase surrounding the BCG has remained stable in the past 9~Gyr of cluster growth. As shown in Fig.~\ref{fig:high_Pc_Lc}, all CC clusters with a significant AGN detection in the BCG present powerful radio-mode feedback ($P_{\mathrm{cav}} > 10^{44}~\mathrm{erg\,s^{-1}}$). This suggests that the onset of radio-mode feedback took place at an early stage ($z \gtrsim 1.5$) of cluster formation.\\
% On average: same amount of energy in cooling and feedback
\indent Moreover, Fig.~\ref{fig:redshift_evol} shows that AGN mechanical feedback is a dominant heating source balancing cooling in cluster cores as $P_{\mathrm{cav}} / L_{\mathrm{cool}} \sim 0.4$ at all redshifts.  The fact that this ratio is not compatible with $1$ at all redshifts does not imply that other significant feedback mechanisms are required to avoid runaway cooling in cluster cores.  A sample average $P_{\mathrm{cav}} / L_{\mathrm{cool}} < 1$ may indeed imply that the AGN duty cycle is lower than 50\% and that the feeding timescale is longer than the feedback one. The hint for a slight negative slope in Fig.~\ref{fig:redshift_evol} is most probably due to the fact that we include upper limits in the analysis of the redshift evolution of $P_{\mathrm{cav}} / L_{\mathrm{cool}}$ while non-detections were not included in previous works.\\
% Implications for the feedback / cooling equilibrium in a progenitor sample
\indent Radio-loud AGN have been shown to have marginal impact on SZ cluster detection with SPT \citep{ble20}.  Assuming the ratio between the number of clusters hosting a radio-loud AGN above a given luminosity threshold and the total number of clusters in our SPT sample is representative of the AGN duty cycle of energy injection $D_{\mathrm{AGN}}$, we find that $D_{\mathrm{AGN}}(z \le 0.72) = 0.5 \pm 0.1$ and $D_{\mathrm{AGN}}(z > 0.72) = 0.4 \pm 0.1$ where we take into account the binomial uncertainties.  This suggests that AGN duty cycles do not evolve significantly with redshift up to $z{\sim}1.5$. While unusually high cooling flows and star formation rates can be observed in individual systems at high redshift (\emph{e.g.} the Phoenix cluster \cite{mcd19}), this work supports a scenario in which radio-mode feedback is able to suppress most of the ICM cooling since the transition between protoclusters and clusters \citep{mul15}.

\subsection{The case of SPT-CLJ2245-6206}\label{subsec:highest_Pcav_Lcool}

We discuss the results obtained in the particular case of SPT-CLJ2245-6206 at $z = 0.58$. This cluster is characterized by $P_{\mathrm{cav}} / L_{\mathrm{cool}} \simeq 91$, \emph{i.e.} the highest feedback response to cooling observed in our sample. This cluster displays the second highest value of $P_{\mathrm{cav}}$ in Fig.~\ref{fig:high_Pc_Lc} and its cooling luminosity is an order of magnitude lower than the one observed in most CC in this sample. The particularity of this cluster is that it displays a clear bimodal morphology (see Fig.~\ref{fig:spt_cl_2245}). The main halo (south) hosting a radio loud AGN in the BCG (magenta circle) presents a rather spherical morphology and is merging with a second sub-halo (north). Although the central density of the main halo is high enough to categorize it as a CC, it presents hints of disturbance that may have been caused by the passage of a shock front induced by the on-going merger event. This would explain the lower cooling luminosity measured at the core of this cluster. This system may be considered as a transition state between the CC and NCC sub-samples studied in this paper.

\subsection{Feedback in non-cool core clusters}\label{subsec:non_cc_clusters}

We investigate how the results presented in this work change if we include clusters categorized as non-cool cores (NCC) in the analysis. We stress that the radio power estimates given in Table~\ref{tab:all_res_NCC} are associated with AGN detections at the location of the BCG.  If no radio AGN is detected in the cluster field or if it is located at $>5~$arcsec from the location of the BCG we provide an upper limit on the radio power.\\
\indent As shown in Fig.~\ref{fig:redshift_evol}, we find that the $P_{\mathrm{cav}} / L_{\mathrm{cool}}$ ratios estimated in NCC clusters are on average higher than the ones found in CC clusters. However, adding the NCC clusters to the fit presented in \textsection \ref{sec:results} does not change the constraints on the slope significantly but rather increases the intrinsic scatter of the relation by 18\%. As shown in Fig.~\ref{fig:high_Pc_Lc}, the distributions of $P_{\mathrm{cav}}$ estimates in CC and NCC are very similar. The high $P_{\mathrm{cav}} / L_{\mathrm{cool}}$ ratios found in NCC are thus driven by the lower cooling luminosities measured on average in these clusters. We show the Magellan/Megacam $r$ band image of the SPT-CLJ0542-4100 cluster along with its \chandra\ and ATCA maps in the right panel of Fig.~\ref{fig:high_Pc_Lc} as a representative example of a NCC cluster with a strong radio AGN detection at the location of the BCG. We propose the following interpretation to explain these observations.\\
\indent As all NCC clusters with a significant radio AGN detection in this sample display a disturbed morphology, recent merging events are probably the cause of the core disturbance.  The transition between CC and NCC has been shown to occur on very short timescales compared to the central cooling time in NCC clusters \citep{ros11}.  If the AGN feedback timescale is much longer than the transition between CC and NCC, observing a BCG at the center of a NCC cluster with strong radio emission can be quite likely.  We may therefore be observing the few NCC clusters displaying a high $P_{\mathrm{cav}} / L_{\mathrm{cool}}$ ratio in a state during which the radio emission is still on-going but the AGN feeding has been reduced by stirring and mixing turbulence that perturbs the dense CC.  The radio power that we measure would then be comparable with that found in CC,  but the X-ray luminosity measured in the cooling radius would be much lower. In this scenario, we expect the radio power to decrease once the feedback response from previous feeding ends. At this point, a dense core would form again at the BCG location and the gas cooling time would drop below the threshold that enables condensation to occur \citep{gas20}. This scenario is also supported by the properties of SPT-CLJ2245-6206 discussed in \textsection \ref{subsec:highest_Pcav_Lcool}. Deeper X-ray data combined with dedicated simulations would however be necessary to test this scenario.

\section{Summary}\label{sec:conclu}

We report the first characterization of the feedback/cooling equilibrium in the core of progenitor-selected clusters at $0.4 < z < 1.4$. Unlike previous studies focusing on the detection of X-ray cavities to estimate the AGN feedback response to gas cooling, we use dedicated ATCA radio observations in the $[2-9]$~GHz band in order to estimate the radio power at 1.4~GHz. This allows us to estimate the AGN cavity power $P_{\mathrm{cav}}$ from the use of a previously calibrated scaling relation between AGN radio and cavity powers whose evolution is assumed to be negligible with redshift. The joint analysis of \chandra\ X-ray and SPT SZ observations of the high-redshift clusters in our sample enables both the estimation of their X-ray luminosity within the cooling radius, $L_{\mathrm{cool}}$, and its comparison with the estimated AGN cavity power. We find that the $P_{\mathrm{cav}} / L_{\mathrm{cool}}$ ratios estimated in these clusters are compatible with the ones found at low redshift. We jointly fit our $P_{\mathrm{cav}} / L_{\mathrm{cool}}$ estimates with the ones obtained in previous studies in order to constrain the redshift evolution of the feedback/cooling equilibrium. We find that this work reduces the uncertainty on the slope of this relation by a factor of 2.3. The latter is compatible with 0 which suggests that radio-mode feedback has balanced gas cooling in the BCG for more than 9~Gyr. This work highlights the importance of joint multi-wavelength analyses to push the investigation of AGN feedback towards higher redshift before the next generation X-ray observatories such as \emph{Athena} \citep{bar20} come into play.

\section*{Acknowledgements}

\small{Support for this work was provided by the National Aeronautics and Space Administration through SAO Award Number SV2-82023 and Chandra Award Number GO9-20117A issued by the Chandra X-Ray Observatory Center, which is operated by the Smithsonian Astrophysical Observatory for and on behalf of NASA under contract NAS8-03060. This work was performed in the context of the South-Pole Telescope scientific program. SPT is supported by the National Science Foundation through grants PLR-1248097,  OPP-1852617,  AST-1814719,  and AST-2109035. Partial support is also provided by the NSF Physics Frontier Center grant PHY-0114422 to the Kavli Institute of Cosmological Physics at the University of Chicago, the Kavli Foundation and the Gordon and Betty Moore Foundation grant GBMF 947 to the University of Chicago. Argonne National Laboratory's work was supported by the U.S. Department of Energy, Office of High Energy Physics, under contract DE-AC02-06CH11357. B.B. is supported by the Fermi Research Alliance LLC under contract no. De-AC02-07CH11359 with the U.S. Department of Energy.  GM acknowledges funding from the European Union's Horizon 2020 research and innovation programme under the Marie Sk\l{}odowska-Curie grant agreement No MARACHAS - DLV-896778. The Australia Telescope Compact Array is part of the Australia Telescope National Facility (\url{https://ror.org/05qajvd42}) which is funded by the Australian Government for operation as a National Facility managed by CSIRO. We acknowledge the Gomeroi people as the traditional owners of the Observatory site. CR acknowledges support from the Australian Research Council's Discovery Projects scheme (DP200101068). }

\begin{table*}
\caption{{\footnotesize Sample properties of the 27 selected CC clusters -- (1) name; (2) redshift; (3) equatorial coordinates of the radio source if we find significant radio signal in the ATCA data. If no radio source is detected, we use the location of the X-ray peak found in the \chandra\ data; (4) radius at which the cooling time is estimated to be 7.7~Gyr; (5) X-ray luminosity measured within the cooling radius; (6) radio power estimated at 1.4~GHz; (7) expected cavity power given the measured radio power and the scaling relation from \cite{cav10}. }}
\renewcommand{\arraystretch}{1.05}
\begin{tabular}{ccccccc}
\hline
\hline
(1) & (2) & (3) & (4) & (5) & (6) & (7)\\
Name & $z$ & $[\mathrm{R.A.,\,Dec.}]$ & $r_{\mathrm{cool}}$ & $L_X(r<r_{\mathrm{cool}})$ & $P_{\mathrm{1.4~GHz}}$ & $\hat{P}_{\mathrm{cav}}$ \\
 & & $[\mathrm{deg,\,deg}]$ & $\mathrm{[kpc]}$ & $\mathrm{[10^{44}~erg\,s^{-1}]}$ & $\mathrm{[10^{40}~erg\,s^{-1}]}$ & $\mathrm{[10^{44}~erg\,s^{-1}]}$ \\
\hline
SPT-CLJ0509-5342 & $0.46$ & $[77.339,-53.703]$ & $112.0^{+8.0}_{-7.0}$ & $2.41\pm 0.1$ & $1.83\pm 0.11$ & $1.3^{+1.0}_{-0.4}$ \\ 
SPT-CLJ0334-4659 & $0.49$ & $[53.546,-46.996]$ & $142.0^{+4.0}_{-4.0}$ & $4.7\pm 0.08$ & $8.07\pm 0.71$ & $3.9^{+5.0}_{-1.6}$ \\ 
SPT-CLJ0346-5439 & $0.53$ & $[56.731,-54.649]$ & $125.0^{+4.0}_{-5.0}$ & $2.2\pm 0.09$ & $21.0\pm 1.22$ & $8.0^{+13.3}_{-3.9}$ \\ 
SPT-CLJ2245-6206 & $0.58$ & $[341.259,-62.128]$ & $56.0^{+5.0}_{-4.0}$ & $0.31\pm 0.04$ & $114.51\pm 3.57$ & $28.5^{+84.1}_{-15.2}$ \\ 
SPT-CLJ2331-5051 & $0.58$ & $[352.962,-50.865]$ & $139.0^{+1.0}_{-3.0}$ & $5.82\pm 0.02$ & $14.04\pm 1.41$ & $5.9^{+8.9}_{-2.5}$ \\ 
SPT-CLJ2232-5959 & $0.59$ & $[338.141,-59.998]$ & $137.0^{+6.0}_{-6.0}$ & $4.78\pm 0.09$ & $<0.33$ & $<0.4$ \\ 
SPT-CLJ0033-6326 & $0.6$ & $[8.471,-63.445]$ & $94.0^{+4.0}_{-3.0}$ & $2.01\pm 0.02$ & $< 0.67$ & $< 0.6$ \\ 
SPT-CLJ0243-5930 & $0.64$ & $[40.863,-59.517]$ & $118.0^{+5.0}_{-5.0}$ & $3.41\pm 0.17$ & $<0.33$ & $<0.4$ \\ 
SPT-CLJ2222-4834 & $0.65$ & $[335.711,-48.576]$ & $124.0^{+10.0}_{-9.0}$ & $3.44\pm 0.18$ & $1.39\pm 0.09$ & $1.0^{+0.7}_{-0.4}$ \\ 
SPT-CLJ0352-5647 & $0.67$ & $[58.24,-56.797]$ & $114.0^{+6.0}_{-5.0}$ & $1.98\pm 0.15$ & $<0.38$ & $<0.4$ \\ 
SPT-CLJ0102-4603 & $0.72$ & $[15.678,-46.071]$ & $84.0^{+6.0}_{-7.0}$ & $0.55\pm 0.09$ & $<0.44$ & $<0.4$ \\ 
SPT-CLJ2329-5831 & $0.72$ & $[352.475,-58.53]$ & $128.0^{+21.0}_{-22.0}$ & $2.22\pm 0.32$ & $<1.23$ & $<0.9$ \\ 
SPT-CLJ2043-5035 & $0.72$ & $[310.823,-50.592]$ & $169.0^{+2.0}_{-2.0}$ & $12.02\pm 0.01$ & $8.8\pm 0.44$ & $4.2^{+5.2}_{-1.8}$ \\ 
SPT-CLJ2301-4023 & $0.73$ & $[345.471,-40.385]$ & $118.0^{+5.0}_{-5.0}$ & $3.42\pm 0.04$ & $<0.5$ & $<0.5$ \\ 
SPT-CLJ2352-4657 & $0.73$ & $[358.068,-46.96]$ & $105.0^{+50.0}_{-24.0}$ & $1.12\pm 0.3$ & $<0.46$ & $<0.5$ \\ 
SPT-CLJ0406-4805 & $0.74$ & $[61.73,-48.082]$ & $123.0^{+13.0}_{-11.0}$ & $2.07\pm 0.21$ & $< 0.73$ & $<0.6$ \\ 
SPT-CLJ2320-5233 & $0.76$ & $[350.121,-52.563]$ & $67.0^{+19.0}_{-16.0}$ & $0.45\pm 0.14$ & $<1.72$ & $<1.2$ \\ 
SPT-CLJ0528-5300 & $0.77$ & $[82.022,-52.998]$ & $94.0^{+95.0}_{-91.0}$ & $0.78\pm 0.09$ & $137.16\pm 8.6$ & $32.6^{+118.5}_{-18.0}$ \\ 
SPT-CLJ2359-5010 & $0.77$ & $[359.928,-50.167]$ & $82.0^{+23.0}_{-17.0}$ & $0.6\pm 0.2$ & $< 1.84$ & $< 1.3$ \\ 
SPT-CLJ0058-6145 & $0.83$ & $[14.588,-61.767]$ & $110.0^{+105}_{-90.0}$ & $1.6\pm 0.12$ & $20.8\pm 1.25$ & $7.9^{+15.3}_{-3.7}$ \\
SPT-CLJ2357-5421 & $0.92$ & $[359.264,-54.364]$ & $221.0^{+34.0}_{-74.0}$ & $3.23 \pm 1.10$ & $<3.35$ & $<2.0$ \\ 
SPT-CLJ2355-5850 & $0.97$ & $[358.9575,-58.850]$ & $129.0^{+32.0}_{-29.0}$ & $1.84 \pm 0.51$ & $<2.57$ & $<1.6$ \\ 
SPT-CLJ2335-5434 & $1.03$ & $[353.882,-54.587]$ & $104.0^{+15.0}_{-10.0}$ & $2.91\pm 0.29$ & $<3.08$ & $<1.9$ \\ 
SPT-CLJ2334-5308 & $1.2$ & $[353.516,-53.141]$ & $126.0^{+17.0}_{-13.0}$ & $4.14\pm 0.73$ & $12.91\pm 1.23$ & $5.5^{+7.5}_{-2.5}$ \\ 
SPT-CLJ2336-5252 & $1.22$ & $[354.081,-52.873]$ & $130.0^{+26.0}_{-28.0}$ & $3.18\pm 0.96$ & $<8.77$ & $<4.1$ \\ 
SPT-CLJ2323-5752 & $1.3$ & $[350.882,-57.881]$ & $195.0^{+45.0}_{-59.0}$ & $4.54\pm 2.7$ & $7.07\pm 0.75$ & $3.5^{+4.1}_{-1.4}$ \\ 
SPT-CLJ2355-5514 & $1.32$ & $[358.874,-55.246]$ & $100.0^{+30.0}_{-33.0}$ & $1.84\pm 1.03$ & $14.09\pm 1.31$ & $5.9^{+9.7}_{-2.8}$ \\ 
\hline
\end{tabular}
\label{tab:all_res}
\end{table*}

\begin{table*}
\caption{{\footnotesize Same as Table~\ref{tab:all_res} for the 21 selected non-cool core clusters. These clusters are only considered in \textsection \ref{subsec:non_cc_clusters}.}}
\renewcommand{\arraystretch}{1.05}
\begin{tabular}{ccccccc}
\hline
\hline
(1) & (2) & (3) & (4) & (5) & (6) & (7)\\
Name & $z$ & $[\mathrm{R.A.,\,Dec.}]$ & $r_{\mathrm{cool}}$ & $L_X(r<r_{\mathrm{cool}})$ & $P_{\mathrm{1.4~GHz}}$ & $\hat{P}_{\mathrm{cav}}$ \\
 & & $[\mathrm{deg,\,deg}]$ & $\mathrm{[kpc]}$ & $\mathrm{[10^{44}~erg\,s^{-1}]}$ & $\mathrm{[10^{40}~erg\,s^{-1}]}$ & $\mathrm{[10^{44}~erg\,s^{-1}]}$ \\
\hline
SPT-CLJ0217-5245 & $0.34$ & $[34.312,-52.76]$ & $49.0^{+5.0}_{-4.0}$ & $0.11\pm 0.01$ & $< 0.1$ & $< 0.2$ \\ 
SPT-CLJ0655-5234 & $0.47$ & $[103.972,-52.57]$ & $39.0^{+9.0}_{-7.0}$ & $0.12\pm 0.01$ & $<0.16$ & $<0.2$ \\ 
SPT-CLJ0200-4852 & $0.5$ & $[30.142,-48.871]$ & $111.0^{+6.0}_{-5.0}$ & $1.5\pm 0.11$ & $< 0.27$ & $< 0.3$ \\ 
SPT-CLJ2306-6505 & $0.53$ & $[346.723,-65.088]$ & $31.0^{+5.0}_{-6.0}$ & $0.04\pm 0.02$ & $< 0.39$ & $< 0.4$ \\
SPT-CLJ2335-4544 & $0.55$ & $[353.784,-45.741]$ & $46.0^{+10.0}_{-7.0}$ & $0.46\pm 0.03$ & $<0.23$ & $<0.3$ \\ 
SPT-CLJ0307-5042 & $0.55$ & $[46.961,-50.701]$ & $66.0^{+9.0}_{-8.0}$ & $0.71\pm 0.05$ & $< 0.25$ & $< 0.3$ \\ 
SPT-CLJ0456-5116 & $0.56$ & $[74.118,-51.278]$ & $55.0^{+13.0}_{-10.0}$ & $0.39\pm 0.03$ & $36.63\pm 1.92$ & $12.1^{+26.8}_{-6.1}$ \\ 
SPT-CLJ2148-6116 & $0.57$ & $[327.179,-61.28]$ & $60.0^{+4.0}_{-3.0}$ & $0.42\pm 0.03$ & $9.28\pm 0.52$ & $4.3^{+5.0}_{-1.9}$ \\ 
SPT-CLJ0256-5617 & $0.58$ & $[44.106,-56.298]$ & $53.0^{+14.0}_{-11.0}$ & $0.39\pm 0.12$ & $4.48\pm 0.29$ & $2.5^{+2.6}_{-0.9}$ \\ 
SPT-CLJ0307-6225 & $0.58$ & $[46.82,-62.447]$ & $55.0^{+8.0}_{-7.0}$ & $0.42\pm 0.03$ & $20.19\pm 1.08$ & $7.7^{+13.5}_{-3.4}$ \\ 
SPT-CLJ0123-4821 & $0.62$ & $[20.8,-48.357]$ & $49.0^{+8.0}_{-6.0}$ & $0.24\pm 0.02$ & $<0.29$ & $<0.3$ \\ 
SPT-CLJ0542-4100 & $0.64$ & $[85.708,-41.0]$ & $68.0^{+4.0}_{-4.0}$ & $0.74\pm 0.06$ & $58.57\pm 2.36$ & $17.2^{+40.8}_{-8.9}$ \\ 
SPT-CLJ2218-4519 & $0.65$ & $[334.749,-45.316]$ & $57.0^{+15.0}_{-11.0}$ & $0.39\pm 0.03$ & $40.21\pm 2.17$ & $13.0^{+25.6}_{-6.5}$ \\ 
SPT-CLJ0310-4647 & $0.71$ & $[47.634,-46.785]$ & $102.0^{+10.0}_{-8.0}$ & $2.2\pm 0.2$ & $< 0.74$ & $< 0.7$ \\ 
SPT-CLJ0324-6236 & $0.73$ & $[51.051,-62.599]$ & $116.0^{+4.0}_{-5.0}$ & $2.73\pm 0.12$ & $< 0.48$ & $< 0.5$ \\ 
SPT-CLJ2328-5533 & $0.77$ & $[352.181,-55.567]$ & $162.0^{+25.0}_{-35.0}$ & $3.85\pm 0.29$ & $<2.56$ & $<1.6$ \\ 
SPT-CLJ2343-5024 & $0.88$ & $[355.837,-50.4]$ & $138.0^{+45.0}_{-59.0}$ & $1.62\pm 0.12$ & $19.35\pm 1.67$ & $7.5^{+12.5}_{-3.4}$ \\ 
SPT-CLJ0533-5005 & $0.88$ & $[83.403,-50.1]$ & $52.0^{+60.0}_{-50.0}$ & $0.16\pm 0.01$ & $27.38\pm 2.44$ & $9.7^{+21.5}_{-4.7}$ \\ 
SPT-CLJ2304-5718 & $0.9$ & $[346.107,-57.306]$ & $27.0^{+16.0}_{-14.0}$ & $0.14\pm 0.01$ & $<3.06$ & $<1.9$ \\ 
SPT-CLJ2311-5820 & $0.93$ & $[347.991,-58.343]$ & $185.0^{+62.0}_{-61.0}$ & $2.62\pm 0.2$ & $26.22\pm 2.71$ & $9.4^{+16.8}_{-4.5}$ \\ 
SPT-CLJ2325-5116 & $0.94$ & $[351.384,-51.285]$ & $63.0^{+34.0}_{-36.0}$ & $0.4\pm 0.03$ & $<2.26$ & $<1.5$ \\ 
\hline
\end{tabular}
\label{tab:all_res_NCC}
\end{table*}

\end{document}